\documentstyle[twocolumn,epsf]{jpsj}

\def\runtitle{Modified Perturbation Theory Applied to the Transport through a Quantum Dot}
\def\runauthor{Osamu {\sc Takagi} and Tetsuro {\sc Saso}}

\title
{Modified Perturbation Theory Applied to Kondo-Type Transport through a Quantum Dot under a Magnetic Field
}

\author
{
Osamu {\sc Takagi}\footnote{E-mail: takagi@krishna.th.phy.saitama-u.ac.jp} and Tetsuro {\sc Saso}\footnote{E-mail: saso@phy.saitama-u.ac.jp} 
}

\inst
{
Department of  Physics, Saitama University, Urawa, Saitama 338-8570
}

\recdate
{
\today
}

\abst
{
Linear conductance through a quantum dot is calculated under a finite magnetic field
 using the modified perturbation theory. The method is based on the second-order
 perturbation theory with respect to the Coulomb repulsion, but the self-energy
 is modified to reproduce the correct atomic limit and to fulfill the Friedel sum
 rule exactly. Although this method is applicable only to zero temperature in a strict
 sense, it is approximately extended to finite temperatures. It is found that the
 conductance near electron-hole symmetry is suppressed by the application of the magnetic field
 at low temperatures. Positive magnetoconductance is observed in the case of large
 electron-hole asymmetry.
}

\kword{quantum  dot, Anderson model, Kondo  effect, perturbation theory, conductance, magnetic field}

\begin{document}
\maketitle

To date, numerous theoretical as well as experimental studies have been carried out on the
 tunneling transport of an electron through
 a quantum dot connected to the leads.
\cite{Book-Single-1992,Book-Meso-1997}
The combination of single electron charging and energy level quantization causes the Coulomb
 blockade or Coulomb oscillation phenomena, which reflects the particle character of an
 electron.
Furthermore, a dot with an odd number of electrons resembles a magnetic impurity coupled to
 the conduction electrons in a metal, and hence, the Kondo-type phenomenon, which reflects
 the wave character of electrons, has been predicted by several researchers
\cite{PRL61-1768-1988,JETP47-452-1988,JPSJ60-3222-1991} and several observations  of Kondo-assisted tunneling were reported recently.
\cite{PRL72-3401-1994,nature391-156-1998,science281-540-1998,condmat-9807233}
The Kondo effect in a quantum dot system has brought up new and interesting issues for
 physics, $e.g.$, the tunable Kondo effect or the nonequilibrium Kondo effect. 

Several theoretical methods have been devised to explain Kondo-type transport through
 a quantum dot using the impurity Anderson model: second-order perturbation theory
 (SOPT),\cite{PRL-3720-1991,PRB-7046-1992,PRB-11622-1996} modified perturbation
 theory,\cite{LevyYeyati93,condmat-9811087,JPSJ-takagi-1998} slave boson method,
\cite{JPSJ-1860-1998} noncrossing approximation (NCA),
\cite{PRL70-2601-1993,PRB49-11040-1994,superlattice22-75-1997} quantum Monte Carlo
 (QMC)\cite{PRB51-4715-1995} and numerical renormalization group (NRG).
\cite{JPSP67-2444-1998} 
Yeyati $et$ $al$.\cite{LevyYeyati93} proposed an interpolative scheme that reproduces
 the correct atomic limit in addition to the weak correlation limit, and applied it
 to transport through a dot.
We have reinvestigated their method and modified it to create a more natural scheme by
 introducing the effective energy level within a dot to fulfill the Friedel sum rule
 exactly.\cite{JPSJ-takagi-1998} 
The main features of a dot were successfully calculated by this method. 
Our scheme, however, was applicable only to zero temperature in the strict sense of the Friedel
 sum formula. 

In this letter, we approximately extend our previous study to finite temperatures,
 and show the temperature and external magnetic field dependence of the linear conductance
 through a quantum dot in the Kondo regime. We also investigate the  spin susceptibility
 and specific heat of the dot, and compare them with the Bethe Ansatz solution of the
 Anderson model to confirm the accuracy of our method.
As a result of numerical calculation, the linear conductance through a dot decreases
 largely with increasing temperature near electron-hole symmetry.
 This result is consistent with those of previous theoretical calculations
\cite{JPSJ60-3222-1991,JPSP67-2444-1998,condmat-9811087} and agrees with experimental
 results.\cite{nature391-156-1998,science281-540-1998,condmat-9807233}.
As regards the magnetic field dependence, we demonstrate the reduction of
 the conductance near electron-hole symmetry which is similar to the temperature
 dependence. 
Moreover, our calculation predicts the positive magnetoconductance in the electron-hole
 asymmetric case. Although it was already pointed out in our previous paper for a limited
 case,\cite{JPSJ-takagi-1998} the behavior of the conductance is clarified in the present
 letter for a wider range of parameters.   

The Hamiltonian for a quantum dot connected to the leads is written as
\begin{eqnarray}
H=
\sum_{\nu,k,\sigma}
\varepsilon^{\nu}_{k}n^{\nu}_{k,\sigma}+
\sum_{\sigma}
\varepsilon_{0}n_{\sigma}+
Un_{\uparrow}n_{\downarrow}\cr
+\sum_{\nu,k,\sigma}
V^{\nu}_{k}(c_{k,\sigma}^{\nu\dagger} c_{0,\sigma} +c_{0,\sigma}^\dagger c_{k,\sigma}^{\nu}) ,
\end{eqnarray}
where $\varepsilon_{0}$ and $U$  represent an energy level and the Coulomb repulsion in a
 dot. The change of $\varepsilon_{0}$ is equivalent to the change of the gate voltage in
 measurements.
$\varepsilon_{k}^{\nu}$ denotes the conduction electron energy in the lead $\nu$(=$R$ and
 $L$).
$V_{k}^{\nu}$ denotes the coupling between leads and a dot. 
We neglect orbital degeneracy and the $k$ dependence of $V_{k}^{\nu}$.
We assume the quantum limit and strong coupling, therefore, the relation
 $\Delta\varepsilon$ $\gg$ $U$ $>$ $\Gamma$ $\gg$ $T$ should be satisfied, where
 $\Delta\varepsilon$ is the spacing of the discrete level, $\Gamma$ is the resonant
 level width and T is the temperature.

Green's function for the electron in the dot is given by
\begin{equation}
G_{\sigma}(\omega)=
\frac{1}
 {
  \omega-\varepsilon_{0} -Un_{-\sigma}-\Sigma_{\sigma}(\omega)+\mbox{i}\Gamma
 } ,  
\end{equation}
where $n_{\sigma}$ denotes the electron number in the dot and $\Sigma_{\sigma}(\omega)$ is the self-energy to be calculated in terms of the
 zeroth-order Green's function $G^{(0)}_{\sigma}(\omega)=(\omega-\varepsilon_{0}+\mbox{i}\Gamma)^{-1}$, $\Gamma/2 = \Gamma_{\scriptsize L} = \Gamma_{\scriptsize R} = \pi\rho_c(0)V^{2}$.
 $\rho_c(0)$ denotes the density of states (DOS) of conduction electrons at the Fermi level. 
In our modified perturbation  theory,  we  introduce an effective  energy  level
 $\tilde{\varepsilon}_{0\sigma}$ in place of the Hartree-Fock level 
$\varepsilon_{0}+Un_{-\sigma}$ and determine $\tilde{\varepsilon}_{0\sigma}$ to  satisfy
 the Friedel  sum  rule exactly.  
We term  this method  the  self-consistent  energy  level second-order  perturbation
  theory (SCEL-SOPT)\cite{JPSJ-takagi-1998}.
Thus, the effective first-order  Green's function  is  given  by
\begin{equation}
G^{(1)}_{\sigma}(\omega)=
\frac{1}
 {
  \omega-\tilde{\varepsilon}_{0\sigma}+\mbox{i}\Gamma
 }.
\end{equation}
Using this Green's function, we first calculate the ordinary second-order self-energy
 $\Sigma_{\sigma}^{(2)}(\omega)$.
Then, we introduce the modified self-energy which is  correct in the atomic limit
 ( $\Gamma$ $\rightarrow$ 0 , $\rho_{\sigma}(\omega)$ $\rightarrow$ $\delta
 (\omega -\tilde{\varepsilon}_{0\sigma}$) ) as\cite{LevyYeyati93,Martin82}
\begin{equation}
\Sigma_{\sigma}(\omega)=
\frac{\Sigma^{(2)}_{\sigma}(\omega)}{1-B\Sigma^{(2)}_{\sigma}(\omega)},
\end{equation}
where
\begin{equation}
B=
\frac
 {
  U(1-n^{(1)}_{-\sigma})+\varepsilon_{0} -\tilde{\varepsilon}_{0\sigma}
 }
 {
   U^2 n^{(1)}_{-\sigma}(1-n^{(1)}_{-\sigma})
 },
\end{equation}
$n^{(1)}_{\sigma}=n [G^{(1)}_{\sigma}]$ and
\begin{equation}
n[G_{\sigma}]\equiv\int\!\! d\omega f(\omega)(-1/\pi)\mbox{Im}G_{\sigma}(\omega).
\end{equation}
Note that $B$ vanishes in the electron-hole symmetric case ( $\varepsilon_{0}$=$-$$U$/2 ).

Next, we construct the second-order Green's function as
\begin{equation}
G^{(2)}_{\sigma}(\omega)=
\frac{1}
 {
  \omega-\varepsilon_{0} -Un^{(1)}_{-\sigma}-
  \Sigma_{\sigma}(\omega)+\mbox{i}\Gamma
 } .   
\end{equation}
Using this $G^{(2)}_{\sigma}(\omega)$, we calculate the second-order electron number
 $n^{(2)}_{\sigma}=n[G^{(2)}_{\sigma}]$.
Furthermore, $Un^{(1)}_{-\sigma}$ in the denominator of eq. (7) is replaced by
 $Un^{(2)}_{-\sigma}$ to obtain the solutions for $\tilde{\varepsilon}_{0}$ over a wide range.
We calculate $n^{(2)}_{\sigma}=n[G^{(2)}_{\sigma}]$ again and
 $n^{(2)\mbox{\scriptsize FS}}_{\sigma}=n^{\mbox{\scriptsize FS}}[G^{(2)}_{\sigma}]$ from
 the Friedel sum formula
\begin{equation}
n^{\mbox{\scriptsize FS}}[G_{\sigma}]\equiv \frac{1}{2}+\frac{1}{\pi}\tan^{-1}\frac{\mbox{Re}(G_{\sigma}(0)^{-1})}{\mbox{Im}(G_{\sigma}(0)^{-1})}.
\end{equation}
Then, we determine $\tilde{\varepsilon}_{0\sigma}$ so as to  
satisfy the relation $n^{(2)}_{\sigma}=n^{(2)\mbox{\scriptsize FS}}_{\sigma}$
 (Friedel sum rule).
\begin{figure}\vspace{0.5cm}
\epsfxsize=7cm
\centerline{\epsfbox{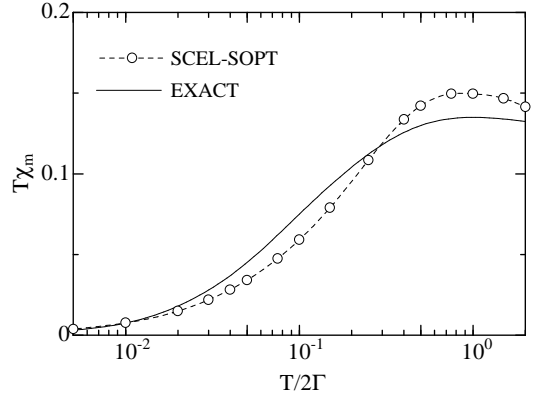}}
\caption{The spin susceptibility multiplied by the temperature is shown  for  $U$=4 ,
 $\Gamma$=1 and $\varepsilon_{0}$=$-$2.  The full line represents the results calculated
 by the Bethe
 Ansatz solution and the circles represent  our calculated results.}
\label{fig:1}
\end{figure}
\begin{figure}\vspace{0.5cm}
\epsfxsize=7cm
\centerline{\epsfbox{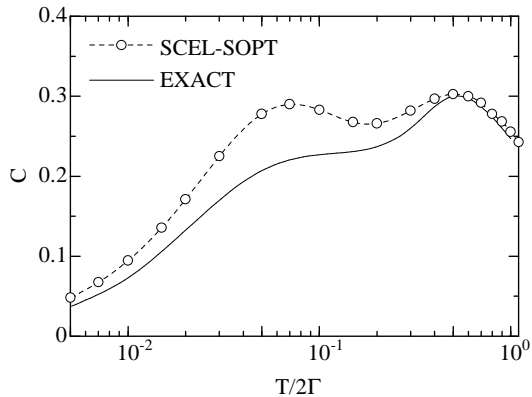}}
\caption{The specific heat is shown  for  $U$=4 , $\Gamma$=1 , $h$=0 and
 $\varepsilon_{0}$=$-$2. The full line represents the results calculated by the Bethe
 Ansatz solution and the circles represent our calculated results.}
\label{fig:2}
\end{figure}
\begin{figure}\vspace{0.5cm}
\epsfxsize=7cm
\centerline{\epsfbox{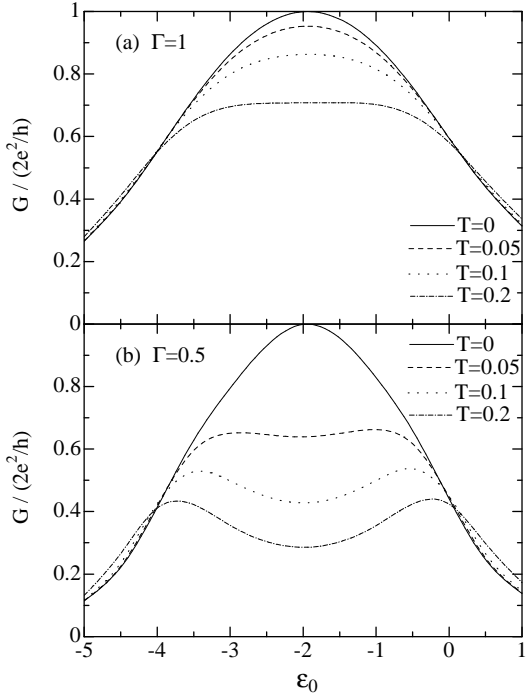}}
\caption{Temperature dependence of conductance is shown for $U$=4 , $h$=0 and $\Gamma$=1(a), 0.5(b).}
\label{fig:3}
\end{figure}
\begin{figure}
\vspace{0.5cm}
\epsfxsize=7cm
\centerline{\epsfbox{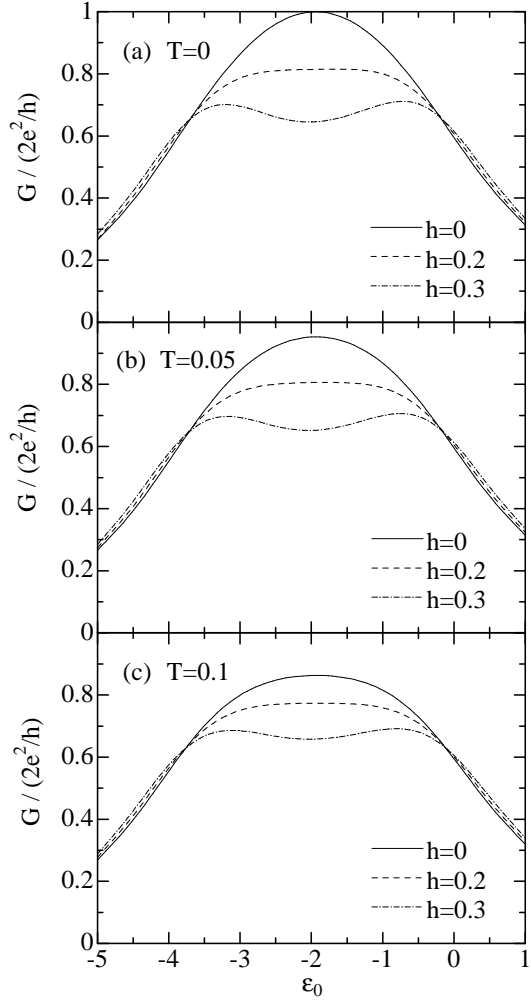}}
\caption{Magnetic field dependence of conductance is shown for $U$=4 , $\Gamma$=1 and $T$=0(a), 0.05(b) and 0.1(c).}
\label{fig:4}
\end{figure}

The present scheme (SCEL-SOPT) is applicable only to zero temperature, since the Friedel
 sum formula is only valid in that case. 
However, we extend this scheme approximately to investigate the properties of a dot at
 finite temperatures by using $\tilde\varepsilon_{0\sigma}$ determined at zero
 temperature.\cite{Krishuna}

To confirm the accuracy of our approximation, we compare several properties with those of
 the Bethe Ansatz solution to the impurity Anderson model.
Figure 1 shows $T\chi_{m}$ where $\chi_{m}$ is the spin susceptibility for U=4, $\Gamma$=1
 and $\varepsilon_{0}$=$-$2. 
The unit of energy here and throughout the present letter should be meV for
 typical quantum dots.
The full line represents the results of the exact solution\cite{Okiji-1157-1983} and the circles
 represent those of our calculations.
 At high temperatures, the exact solution gradually approaches  $1/8$. Our result shows
 qualitative agreement with the exact solution over a wide range of temperatures. 
Figure 2 indicates the specific heat $C=\mbox{d}E_{\mbox{\scriptsize{tot}}}/\mbox{d}T$,
 where $E_{\mbox{\scriptsize{tot}}}$ is the total energy calculated from 
\begin{eqnarray}
E_{\mbox{\scriptsize{tot}}}=
\sum_{\sigma}\int_{-\infty}^{\infty}
d\omega f(\omega) (-\frac{1}{\pi}) \ \ \ \ \ \ \ \ \ \ \ \ \ \ \ 
\cr
\times \mbox{Im}[\{ \omega -\frac{1}{2} (Un_{-\sigma}^{(2)}+\Sigma_{\sigma}(\omega)) \}G^{(2)}_{\sigma}(\omega)].
\end{eqnarray}
Here, $\Sigma_{\sigma}(\omega)$ is the modified self-energy given by eq. (4).
The global feature of $C(T)$ is well reproduced by the present calculation.

As we have assumed a single energy level $\varepsilon_{0}$ in a quantum dot, the current takes the form\cite{PRL-2512-1992}
\begin{equation}
I=\frac{2e}{\hbar}\sum_{\sigma}\int d\omega
\{f_{\scriptsize L}(\omega )-f_{\scriptsize R}(\omega )\}
\Gamma^{'}\rho_{\sigma}(\omega ),
\end{equation}
where $\Gamma^{'} =\Gamma_{\scriptsize L}\Gamma_{\scriptsize R}/
(\Gamma_{\scriptsize L}+\Gamma_{\scriptsize R})$ 
and $f_{\nu}(\omega)$ is the Fermi function of the lead $\nu$. We utilize eq. (10)
 to obtain the differential conductance ($dI/dV$) at zero bias voltage which is equivalent
 to the linear conductance, where the bias voltage is given by the difference of the chemical
 potential between the left and right leads as $V$=$\mu_{L}$$-$$\mu_{R}$. 
We can use an equilibrium DOS for $\rho_{\sigma}(\omega)$ in eq. (10) to calculate the 
transport properties in a linear response regime. 

Figure 3 shows the temperature dependence of the linear conductance through a quantum dot
 for $U$=4, $h$=0 and $\Gamma$=1(a), 0.5(b) as a function of the gate voltage, where
 $h=\frac{1}{2}g\mu_{B}H$.  
The full, broken, dotted and  dash-dotted lines represent the results for $T$=0, 0.05, 0.1 and
 0.2, respectively.
At low temperatures, the Kondo peak of DOS emerges at the Fermi level in the symmetric
 case, and hence, the conductance takes the value 2$e^{2}/h$ due to the Kondo-assisted
 tunneling in that case.
In the case of electron-hole asymmetry, the Kondo peak shifts from the Fermi level
 and the conductance is reduced.
As the Kondo effect does not appear in the extreme asymmetric case, $\varepsilon_{0}<$$-$4
 or $\varepsilon_{0}>$0 for example, the conductance does not
 depend on the temperature. 
In Fig. 3(b), the conductance exhibits two peaks for $T$=0.2 which is nothing but the
 Coulomb
 oscillation. The peaks appear when the discrete energy level of a quantum dot
 $\varepsilon_{0}$ or $\varepsilon_{0}+U$ agrees with the Fermi level. These two peaks
 are not seen clearly in Fig. 3(a) since the width of the resonant level $\Gamma$ is large.
These temperature dependencies of the conductance were already obtained theoretically by
 other groups\cite{JPSJ60-3222-1991,JPSP67-2444-1998,condmat-9811087}, and our numerical
 calculation is consistent with their results.
 Recent experiments\cite{nature391-156-1998,science281-540-1998,condmat-9807233} have
 revealed Kondo-assisted tunneling at low temperatures that demonstrates qualitative
 agreement with our results.

Figure 4 shows the magnetic field dependence of the conductance for $U$=4 and $\Gamma$=1.
The full, broken and dash-dotted lines represent the solution for $h$=0, 0.2 and 0.3,
 respectively. We have included an external magnetic field $H$ by shifting the dot
 level as $\varepsilon_{0}\pm h$ for up and down spins.
The chemical potential $\mu_{\nu}$'s are assumed to be unchanged by the field.
Under the magnetic field, the Kondo peak shifts to the right and left for up and down
 spins, respectively,
 by Zeeman energy. Then the DOS at the Fermi level, which determines the transport properties,
 decreases, and hence, the conductance near electron-hole symmetry is reduced. 
This behavior appears to be consistent with that recently observed in experimens.
\cite{nature391-156-1998,
science281-540-1998}
The above feature is similar to the case of reduction in conductance with increasing
 temperature in Fig. 3.
In the extreme asymmetric case, the opposite feature emerges which enhances the
 conductance by the application of the magnetic field, namely, positive magnetoconductance.
As regards the sign of the magnetoconductance, we assume that, in the range of negative
 magnetoconductance, the dot contains an odd number of electrons and the Kondo-like effect
 occurs. Otherwise, there are an even number of electrons and the Kondo effect does
 not occur. 
Furthermore, we observe that increasing the temperature weakens the magnetic field effect.

In conclusion, we have studied the transport properties through a quantum dot in the  Kondo
 regime by the modified perturbation theory (SCEL-SOPT), which modifies the self-energy so as to
 reproduce the correct atomic limit and to fulfill the Friedel sum rule exactly.
 We extended this theory to study the finite temperature properties of a quantum dot.
The accuracy of our scheme was confirmed by comparing the spin susceptibility and the specific
 heat with the exact solution.
It has been demonstrated that our scheme can be used for a wide range of temperatures 
and magnetic field strengths.
As regards the transport properties through a dot, we noted the reduction of the
 conductance near electron-hole symmetry under magnetic field at low temperatures,
 which explains the recent experimental results.
\cite{nature391-156-1998,science281-540-1998} We also showed the positive
 magnetoconductance in the case of asymmetry. Such properties may also be observed
 in the experimental results.
During the preparation of the present manuscript, we discovered a paper\cite{JPSJ-1640-1999}
 which also calculated the magnetic field dependence of the conductance through a
 dot by QMC. 
The result is consistent with that of our calculation near the electron-hole symmetric case.
 However, because of the use of QMC, their calculation does not cover sufficiently low
 temperatures or a wide range of gate voltages. 
Our approach is much simpler with sufficient accuracy and may have wide
 potential applicability to various situations in quantum dot systems.

The computation was performed using FACOM VPP500 in the Supercomputer Center, Institute
 for Solid State Physics, University of Tokyo.
This work is supported by a Grant-in-Aid for Scientific Research No.11640367 from the
 Ministry of Education, Science, Sports and Culture.

\end{document}